\documentclass[preprint,showpacs,preprintnumbers,amsmath,amssymb,prb]{revtex4}

\usepackage{graphicx}
\usepackage{dcolumn}
\usepackage{bm}
\usepackage{longtable}

\begin{document}

\title{Time-dependent density-functional theory
for ultrafast interband excitations}

\author{V.~Turkowski}\email{turkowskiv@missouri.edu}

\author{C.A.~Ullrich} \email{ullrichc@missouri.edu}

 \affiliation{Department of Physics and Astronomy, University of Missouri,
Columbia, MO 65211}

\date{\today}

\begin{abstract}
We formulate a time-dependent density functional theory (TDDFT) in
terms of the density matrix to study ultrafast phenomena in
semiconductor structures. A system of equations for the density
matrix components, which is equivalent to the time-dependent
Kohn-Sham equation, is derived. From this we obtain a TDDFT version
of the semiconductor Bloch equations, where the electronic many-body
effects are taken into account in principle exactly. As an example,
we study the optical response of a three-dimensional two-band
insulator to an external short-time pulsed laser field. We show that
the optical absorption spectrum acquires excitonic features when the
exchange-correlation potential contains a $1/q^{2}$ Coulomb
singularity. A qualitative comparison of the TDDFT optical
absorption spectra with the corresponding results obtained within
the Hartree-Fock approximation is made.
\end{abstract}



\maketitle

\section{Introduction}
\label{Introduction}

Due to the demands of modern electronics, semiconductor devices are
becoming smaller and faster, which means that applied external
fields cause strongly inhomogeneous and nonequilibrium processes in
such systems. A powerful approach to study dynamical properties of
potentially useful devices is to apply short (femto- or picosecond)
electric field pulses and to measure the response of the system.
Recent progress in ultrafast laser pulse experimental techniques
allows one to study the physical processes in such systems with a
very high precision (for a review, see e.g.
Refs.~\onlinecite{Onida,Rossi}). For example, with this technique
one can measure nonequilibrium energy and momentum distributions, or
the dynamics of excited states. Therefore, it is necessary to
develop theoretical tools to describe these experiments, as well as
to understand ultrafast processes in small semiconductor devices in
general.

The theoretical description of ultrafast phenomena triggered by
short-pulse laser fields is a complicated problem due to several
reasons. One of the most difficult tasks is to take into account
many-particle correlation effects properly. An external pulse field
causes the following main effects in the system: i) direct electron
photoemission; ii) inverse electron photoemission, and iii)
absorption processes. The first two phenomena can be described in
terms of free quasi-particles, which makes the problem relatively
simple. A proper description of optical absorption spectra is a much
more complicated task due to quasi-particle correlation effects. In
particular, external laser pulses can create excitons, or coupled
electron-hole pairs. The problem of correctly describing optical
absorption spectra with excitonic features in the case of applied
short-time laser pulses is one of the great challenges in condensed
matter physics.

For small devices in the presence of a short-pulse field typical
time scales are shorter than the Coulomb scattering time (see, for
example, Ref.~\onlinecite{Chemla}), which means that one cannot
treat the Coulomb interaction effects by using a simple Boltzmann
equation approach, where all the Coulomb effects are "hidden" in a
scattering time parameter $\tau$. Similarly, the semiconductor Bloch
equation (SBE) approach, \cite{Haug} based on the Hartree-Fock (HF)
approximation, and other mean-field methods have difficulties under
these circumstances, because of the presence of strong fluctuations.
In principle, Coulomb interaction effects can be taken into account
in a systematic way by using non-equilibrium Green function
techniques, \cite{Kadanoff_Baym,Keldysh} similar to the equilibrium
case. Unfortunately, this technique becomes numerically very
complicated in a strongly nonequilibrium situation, since in this
case the Green functions depend on two or more time arguments, which
puts high requirements on the computer memory size and makes the
numerical analysis very time-consuming.\cite{Onida,Dahlen}

In this paper, we discuss an alternative and potentially very
powerful approach to study these kinds of problems, based on
density-functional theory (DFT),\cite{Kohn} and in particular its
time-dependent generalization
(TDDFT).\cite{Runge_Gross,Gross,TDDFT_book} In this approach,
numerical calculations should be much less time-consuming compared
to the Green's function method. TDDFT has been successfully applied
to describe molecular excitations;\cite{Elliott} however, this
approach has some difficulties in describing extended
systems.\cite{Botti2} It is known that the standard local-density
approximation (LDA) and generalized gradient approximation (GGA) for
the DFT exchange-correlation (xc) potential cannot be applied to
describe some effects beyond the ground state properties in extended
systems, like the energy band gaps and excitonic effects in the
optical absorption spectra. Therefore, in order to apply TDDFT to
study ultrafast processes, and in particular to describe correctly
the optical absorption spectra in such systems, it is necessary to
find suitable xc potentials.

For weak and smooth external fields, such a potential can
constructed by using the many-body Bethe-Salpeter equation (BSE)
approach \cite{Reining,Marini,Botti}. In fact, in this case TDDFT
calculations with an xc potential extracted from the BSE give very
good results for the optical absorption spectra in a set of bulk
semiconductors. Unfortunately, the BSE-TDDFT approach cannot be used
directly to construct an xc potential for excitation with strong
short pulses, since here the linear response theory cannot be
applied, and the equations in time domain depend on many time
variables, which makes numerical solution extremely difficult to
obtain (similar to non-equilibrium Green's functions). Therefore, it
would be extremely useful to find a simple xc potential which will
allow one to use TDDFT straightforwardly to study the optical
response of a system in the time domain. It was shown in
Refs.~\onlinecite{Kim1,Kim2} that for smooth and weak fields such a
potential exists. Namely, the exact-exchange approximation for the
time-dependent optimized effective potential (XX-TDOEP) results in
optical absorption spectra with pronounced excitonic features, by
solving the problem in the frequency domain. Kim and
G\"orling~\cite{Kim1,Kim2} showed that the main reason for this was
the presence of $1/q^2$ Coulomb singularity in the exchange energy
kernel. The XX-TDOEP has recently been extended into the nonlinear,
real-time domain for simple quasi-one-dimensional quantum well
systems.\cite{Wijewardane2}

The purpose of this paper is twofold: 1) we will formulate a general
TDDFT approach to study optical interband excitations in terms of
the Kohn-Sham density matrix. The corresponding system of equations
has the formal simplicity of the SBEs, which allows one to solve it
directly in the time domain, contrary to the many-body Green
function approach, where it is difficult to treat the problem
numerically. At the same time, this approach has a great formal
advantage in comparison with the SBE formalism, since in TDDFT the
Coulomb interaction effects are treated in principle exactly. 2) We
will show that by using simple exchange-only functionals, the
essence of excitonic features can be captured in a relatively simple
manner. As an example, we consider the model of a three-dimensional
two-band bulk insulator for different local xc potentials and show
that its optical absorption spectra contain qualitatively correct
excitonic features, whenever the xc energy kernel has a $1/q^{2}$
singularity.

The paper is organized as follows: We introduce a general TDDFT-SBE
formalism in terms of the Kohn-Sham density matrix in
Section~\ref{Formalism}. In Section~\ref{3D}, we derive the
TDDFT-SBE formalism and in Section~\ref{Results} we apply this
formalism to study the optical absorption spectra for a
three-dimensional two-band model insulator with different xc
potentials and compare the results with HF. Conclusions are
presented in Section~\ref{Conclusions}. Some technical details are
given in Appendix.

We use Hartree atomic units ($e^2=m=\hbar =1$) throughout the paper.

\section{General formalism}
\label{Formalism}

The general DFT Hamiltonian for a many-electron system in a solid
(in the Born-Oppenheimer approximation) can be written in the
following form:
\begin{equation}
{\hat H}=-\frac{{\bf \nabla}^{2}}{2}+V_{nucl}({\bf r})+V_{H}[n]({\bf
r}) +V_{xc}[n]({\bf r}), \label{Hamiltonian}
\end{equation}
where $V_{nucl}({\bf r})$ is the nuclear potential for the electrons
and
\begin{equation}
V_{H}[n]({\bf r})=\int d{\bf r}'\frac{n({\bf r}')}{|{\bf r}-{\bf
r}'|} \label{VH}
\end{equation}
is the Hartree potential, where $n({\bf r})$ is the density of
electrons. All many-body effects beyond Hartree are described by the
scalar xc potential $V_{xc}[n]({\bf r})$. In particular, in the LDA
exchange-only case:
\begin{equation}
V_{xLDA}(n({\bf r})) = -\left(\frac{3}{\pi}
\right)^{1/3}n^{1/3}({\bf r}). \label{VLDA}
\end{equation}

In order to describe the ground-state properties of the system
governed by the Hamiltonian Eq.~(\ref{Hamiltonian}), one solves the
stationary Kohn-Sham (KS) equation
\begin{equation}
{\hat H}({\bf r})\psi_{{\bf k}}^{l_{i}}({\bf r})=\varepsilon_{{\bf
k}}^{l_{i}}\psi_{{\bf k}}^{l_{i}}({\bf r}), \label{Schroedinger}
\end{equation}
and find bands $l_{i}$ with spectra $\varepsilon_{{\bf k}}^{l_{i}}$,
where ${\bf k}$ is the crystal momentum and $l_{i}=v_{i},c_{i}
(i=1,2,...)$ are the labels for the valence ($v$) and conduction
($c$) bands. The electron density can be found self-consistently:
\begin{equation}
n({\bf r})=2\sum_{i,{\bf k}}|\psi_{{\bf k}}^{v_{i}}({\bf
r})|^{2}\theta (\varepsilon_{F}-\varepsilon_{{\bf k}}^{v_{i}}),
\label{n2}
\end{equation}
where $\varepsilon_{F}$ is the Fermi energy and the summation is
performed over the occupied (valence) band states.
Equation~(\ref{n2}) is valid in the case of zero temperature, which
we consider in this paper. Finite temperatures would require
introduction of the Fermi distribution function into Eq.~(\ref{n2}).

In order to study the nonequilibrium case when an external electric
field ${\bf E}(t)$ is switched on at time $t=t_{0}$, the Hamiltonian
Eq.~(\ref{Hamiltonian}) must be modified in the standard way: an
external electromagnetic vector potential ${\bf A}_{ext}({\bf r},t)$
should be added by making the usual substitution ${\bf
\nabla}\rightarrow {\bf \nabla}-(i/c){\bf A}_{ext}({\bf r},t)$ and
by adding a scalar potential term $\varphi_{ext}({\bf r},t)$ to the
Hamiltonian. The electric field is connected with $\varphi_{ext}
({\bf r},t)$ and ${\bf A}_{ext} ({\bf r},t)$ in the following way:
\begin{equation}
{\bf E}({\bf r},t)=-{\bf \nabla}\varphi_{ext}({\bf r},t)
-\frac{1}{c}\frac{\partial {\bf A}_{ext}({\bf r},t)}{\partial t}.
\label{E}
\end{equation}
For an extended system one needs to preserve the periodicity,
therefore the vector external potential must be used. This makes the
problem technically more complicated in comparison to scalar
potentials. However, it can be shown that in situations when the
characteristic field frequency is bigger than the level spacing, one
can work with scalar potentials instead of vector potentials (see,
for example, Ref.~\onlinecite{Schafer}). Therefore, we shall
consider the case when ${\bf A}_{ext}({\bf r},t)=0$, and for
simplicity assume that the electric field is space-independent.
Strictly speaking, for external homogeneous fields one needs to use
current-TDDFT in order to study the response of the
system.\cite{Maitra,vanFaassen} One then gets the macroscopic
current, which allows one to satisfy the periodicity condition. In
this paper, however, we use the usual approximation\cite{Haug} where
the external homogeneous electric field is described by the
following scalar potential:
\begin{equation}
\varphi_{ext}({\bf r},t)=-{\bf E}(t){\bf r}. \label{varphiext}
\end{equation}

Due to the presence of an electric field, the Hamiltonian is
explicitly time-dependent. Moreover, the Hartree and xc potential
terms in the Hamiltonian become also time-dependent. Such a
time-dependent problem is described by the time-dependent KS
equation:
\begin{equation}
i\frac{\partial }{\partial t}\Psi ({\bf r},t)={\hat H}({\bf
r},t)\Psi ({\bf r},t). \label{Schroedinger2}
\end{equation}
In the adiabatic approximation, which we use in this paper, the xc
potential can be formally obtained from the stationary potentials by
assuming that the electron density becomes time-dependent, $n({\bf
r})\rightarrow n({\bf r},t)$. Eq.~(\ref{Schroedinger2}) must be
solved self-consistently together with the corresponding
time-dependent particle number equation (\ref{n2}). Since the
initially occupied states are the valence states $\psi_{{\bf
k}}^{v_{i}}({\bf r})$, described by the band index $v^{i}$ and the
momentum ${\bf k}$, the evolution of the system can be completely
described by the time-evolution of these states, i.e. by finding the
corresponding time-dependent wave functions $\Psi_{{\bf
k}}^{v_{i}}({\bf r},t)$, such that $\Psi_{{\bf k}}^{v_{i}}({\bf
r},t_{0})=\psi_{{\bf k}}^{v_{i}}({\bf r})$, from
Eq.~(\ref{Schroedinger2}) and a time-dependent generalization of
Eq.~(\ref{n2}):
\begin{equation}
n({\bf r},t)=2\sum_{i,{\bf k}}|\Psi_{{\bf k}}^{v_{i}}({\bf
r},t)|^{2}\theta (\varepsilon_{F}-\varepsilon_{{\bf k}}^{v_{i}}).
\label{n3}
\end{equation}

In the following, we express the time-dependent wave functions as
linear combinations of the ground-state wave functions:
\begin{eqnarray}
\Psi_{{\bf k}}^{v_{i}}({\bf r},t)=\sum_{j,{\bf q}}\left[ c_{{\bf
k}{\bf q}}^{v_{i}v_{j}}(t)\psi_{{\bf q}}^{v_{j}}({\bf r}) +c_{{\bf
k}{\bf q}}^{v_{i}c_{j}}(t)\psi_{{\bf q}}^{c_{j}}({\bf r}) \right] ,
\label{psi}
\end{eqnarray}
where $c_{{\bf k}{\bf q}}^{v_{i}l_{j}}(t)$ are momentum and
time-dependent complex coefficients, which satisfy the following
initial condition: $c_{{\bf k}{\bf q}}^{v_{i}l_{j}}(t_{0})=\delta_
{{\bf k}{\bf q}}\delta_{v_{i}l_{j}}e^{-i\varepsilon_{{\bf
k}}^{v_{i}}t_{0}}$. The time evolution of the system can thus be
found by determining the coefficients $c_{{\bf k}{\bf
q}}^{v_{i}v_{j}}(t)$ and $c_{{\bf k}{\bf q}}^{v_{i}c_{j}}(t)$ for
Eq.~(\ref{psi}). However, in order to solve the problem, it is more
convenient to introduce the density matrix: $\rho_{{\bf k};{\bf
q}{\bf p}}^{v_{i};l_{m}{\bar l}_{n}}(t)=c_{{\bf k}{\bf
q}}^{v_{i}l_{m}}(t) \left[ c_{{\bf k}{\bf p}}^{v_{i}{\bar
l}_{n}}(t)\right] ^{*}$, which is useful for defining physical
quantities like occupation of the states and optical transitions
(see next Section). TDDFT in the density-matrix representation is a
method which allows one to solve the problem in principle exactly,
since it is an exact reformulation of the time-dependent Kohn-Sham
formalism. This method was already introduced to study intersubband
processes in quantum wells\cite{Wijewardane} and the electrical
conductivity in dissipative models of molecular devices\cite{Burke}
(see also Ref.~\onlinecite{Berman}, where a similar approach was
discussed). Here we develop a technique which can be applied in more
general cases with a continuous electron spectrum, including
interband transitions in solids.

The density matrix satisfies the following equation of motion:
\begin{equation}
i\frac{\partial}{\partial t}\rho_{{\bf k};{\bf q}{\bf
p}}^{v_{i};l_{m}l_{n}'}(t)=[H(t),\rho ]_{{\bf k};{\bf q}{\bf
p}}^{v_{i};l_{m}l_{n}'} = \sum_{l_{j}'',{\bf q}'}\left[ H_{{\bf
q}{\bf q}'}^{l_{m}l_{j}''}(t)\rho_{{\bf k};{\bf q}'{\bf
p}}^{v_{i};l_{j}''l_{n}'}(t) - \rho_{{\bf k};{\bf q}{\bf
q}'}^{v_{i};l_{m}l_{j}''}(t)H_{{\bf q}'{\bf p}}^{l_{j}''l_{n}'}(t)
 \right] , \label{Liouville}
\end{equation}
where the Hamiltonian matrix elements $H_{{\bf k}{\bf
q}}^{l_{m}{\bar l}_{n}}(t)$ are
\begin{equation}
H_{{\bf k}{\bf q}}^{l_{m}l_{n}'}(t)=\int_{cell} d{\bf r} \psi_{{\bf
k}}^{l_{m}*}({\bf r}) H(t)\psi_{{\bf q}}^{l_{n}'}({\bf r})
=\varepsilon_{{\bf k}}^{l_{m}}\delta_{l_{m}l_{n}'}+{\bf E}(t){\bf
d}_{{\bf k}{\bf q}}^{l_{m}l_{n}'} +V_{H{\bf k}{\bf
q}}^{l_{m}l_{n}'}(t)+V _{xc{\bf k}{\bf q}}^{l_{m}l_{n}'}(t),
\label{Hamiltonianme}
\end{equation}
and the space integration is performed over a unit cell. In
Eq.~(\ref{Hamiltonianme}),
\begin{equation}
{\bf d}_{{\bf k}{\bf q}}^{l_{m}l_{n}'} = \int_{cell} d{\bf r}
\psi_{{\bf k}}^{l_{m}*}({\bf r}) {\bf r}\psi_{{\bf q}}^{l_{n}'}({\bf
r}) \label{dipole}
\end{equation}
are the dipole matrix elements, and $V_{H{\bf k}{\bf q}}^{l_{m}
l_{n}'}(t)$ and $V _{xc{\bf k}{\bf q}}^{l_{m}l_{n}'}(t)$ are the
matrix elements for the difference between the time-dependent and
ground state (at $t\leq t_{0}$) Hartree and xc potentials:
\begin{equation}
V_{H{\bf k}{\bf q}}^{l_{m}l_{n}'}(t) = \int_{cell} d{\bf r}
\psi_{{\bf k}}^{l_{m}*}({\bf r}) \left[ V_{H}[n]({\bf r},t)
-V_{H}[n]({\bf r},t_{0})\right]\psi_{{\bf q}}^{l_{n}'}({\bf
r}),\label{VHlm}
\end{equation}
and similar for $ V_{xc{\bf k}{\bf q}}^{l_{m}l_{n}'}(t)$. The
density matrix satisfies the following initial condition:
\begin{equation}
\rho_{{\bf k};{\bf q}{\bf p}}^{v_{i};l_{m}l_{n}'}(t_{0})=
\delta_{{\bf k}{\bf q}}\delta_{{\bf k}{\bf
p}}\delta_{v_{i}l_{m}}\delta_{v_{i}l_{n}'}, \label{Liouvillet0}
\end{equation}
which corresponds to the situation where all states in the valence
bands are initially occupied. The particle density (\ref{n3}) has
the following form in terms of the density matrix elements:
\begin{equation}
n({\bf r},t)=2\sum_{i,l_{m},l_{n}',{\bf k},{\bf q},{\bf p}}
\rho_{{\bf k},{\bf q},{\bf p}}^{v_{i};l_{m}l_{n}'}(t) \psi_{{\bf
p}}^{l_{n}'*}({\bf r}) \psi_{{\bf q}}^{l_{m}}({\bf r}) \theta
(\varepsilon_{F}-\varepsilon_{{\bf k}}^{v_{i}}). \label{nrho}
\end{equation}
Solution of the Liouville-von Neumann equation (\ref{Liouville})
allows one to study physical properties of the system. In
particular, one can find the polarization as
\begin{equation}
{\bf D}(t)=\sum_{i,l_{m},l_{n}',{\bf k},{\bf q},{\bf p}} \rho_{{\bf
k};{\bf q}{\bf p}}^{v_{i};l_{m}l_{n}'}(t) {\bf d}_{{\bf p}{\bf
q}}^{l_{n}'l_{m}}.
 \label{polarization}
\end{equation}

Below, we shall use this formalism to study the optical response of
a three-dimensional two-band insulator, by solving
Eqs.~(\ref{Liouville}) and (\ref{Liouvillet0}) using different xc
potentials.

\section{Three-dimensional two-band model}
\label{3D}

In general, it is very difficult to find the solution of the density
matrix equation (\ref{Liouville}) and one needs to make some
approximations. For simplicity, we shall consider the optical
absorption spectra of systems composed of one valence and one
conduction band: $l_{1}=v$, $l_{2}=c$. This approximation can be
used if one assumes that the dominant optical transitions in the
system take place from the highest occupied valence band to the
lowest conduction band. Since the main aim of this paper is a {\it
proof of principle} that TDDFT can describe excitonic effects in
semiconductors triggered by short laser pulses, such an
approximation is sufficient. However, for real systems one needs to
take into account the band structure of the materials more
accurately. Another simplification comes from the fact that in the
dipole approximation for the external field, the optical transitions
in the system take place with zero photon momentum. The coefficients
$c$, defined in Eq.~(\ref{psi}), then depend on one momentum
variable only:
\begin{eqnarray}
c_{{\bf k}{\bf q}}^{vv}(t)=\delta_{{\bf k}{\bf q}}c_{{\bf
k}}^{vv}(t), \ \ \ \ c_{{\bf k}{\bf q}}^{vc}(t)=\delta_{{\bf k}{\bf
q}}c_{{\bf k}}^{vc}(t).
 \label{clm}
\end{eqnarray}
The problem is thus reduced to finding the density matrix of rank 2,
with elements $\rho_{{\bf k}}^{vv}(t)$, $\rho_{{\bf
 k}}^{vc}(t)$, $\rho_{{\bf k}}^{cv}(t)$ and $\rho_{{\bf
 k}}^{cc}(t)$, which are functions of momentum and time.
The elements $\rho_{{\bf k}}^{vv}(t)$ and $\rho_{{\bf
 k}}^{cc}(t)$ describe the occupancy of the valence and conduction band states, and
$\rho_{{\bf
 k}}^{vc}(t)$ and $\rho_{{\bf k}}^{cv}(t)$ describe the polarization in
 the system. These four elements are not independent.
 Particle conservation number requires $\rho_{{\bf k}}^{vv}(t)+\rho_{{\bf
 k}}^{cc}(t)=1$, and by definition, $\rho_{{\bf
 k}}^{cv}(t)=\rho_{{\bf k}}^{vc*}(t)$.

From Eq.~(\ref{Liouville}), one can obtain the following system of
equations for
 two independent components $\rho_{{\bf k}}^{vv}(t)$ and $\rho_{{\bf
 k}}^{vc}(t)$:
\begin{eqnarray}
\frac{\partial}{\partial t}\rho_{{\bf k}}^{vv}(t) = &-&2{\rm
Im}\left[ (E(t)d_{\bf k}^{cv}+V_{H{\bf k}}^{cv}+V_{xc{\bf k}}^{cv} )
\rho_{{\bf k}}^{vc}(t)\right] , \label{rhovv}
\\
\frac{\partial}{\partial t}\rho_{{\bf k}}^{vc}(t)=
&-&i[\varepsilon_{\bf k}^{v}-\varepsilon_{\bf k}^{c}] \rho_{{\bf
k}}^{vc}(t) -i [\rho_{{\bf k}}^{cc}(t)-\rho_{{\bf k}}^{vv}(t)]
E(t)d_{\bf k}^{vc}
\nonumber \\
&-i&[\rho_{{\bf k}}^{cc}(t)-\rho_{{\bf k}}^{vv}(t)] (V_{H{\bf
k}}^{vc}(t)+V_{xc{\bf k}}^{vc}(t)) -i[V_{H{\bf k}}^{vv}(t)+V_{xc{\bf
k}}^{vv}(t) -V_{H{\bf k}}^{cc}(t)-V_{xc{\bf k}}^{cc}(t) ]\rho_{{\bf
k}}^{vc}(t).\nonumber \\
\label{rhovc}
\end{eqnarray}
This system represents the TDDFT version of the well-known
SBEs,\cite{Haug} which are used to study the optical properties of
bulk semiconductors subject to external electric fields.
Equations~(\ref{rhovv}), (\ref{rhovc}) have a more general form,
since here the Coulomb interaction effects are taken into account in
principle exactly by means of the matrix elements of $V_{xc}[n]({\bf
r},t)$. As mentioned in the Introduction, this is important in the
case of small systems and sharp pulses, where the characteristic
times in the system are shorter than the Coulomb scattering time,
and where it is difficult to treat the Coulomb interaction effects
properly within the BSE approach.

We shall study the solution of Eqs.~(\ref{rhovv}) and (\ref{rhovc})
for the simple but instructive example of a two-band model on a
cubic lattice. We assume that the solution of
Eq.~(\ref{Schroedinger}) with the Hamiltonian
Eq.~(\ref{Hamiltonian}) gives the following simple dispersions:
\begin{equation}
\varepsilon_{{\bf k}}^{v} =\varepsilon_{0}^{v}+2t^{v} [\cos
(a_{0}k_{x})+\cos (a_{0}k_{y})+\cos (a_{0}k_{z})],
 \label{epskv}
\end{equation}
\begin{equation}
\varepsilon_{{\bf k}}^{c} =\varepsilon_{0}^{c}-2t^{c} [\cos
(a_{0}k_{x})+\cos (a_{0}k_{y})+\cos (a_{0}k_{z})],
 \label{epskc}
\end{equation}
where $a_{0}$ is the lattice constant. For definiteness, we consider
a model of solid hydrogen and assume that the difference of
$\varepsilon_{0}^{c}$ and $\varepsilon_{0}^{v}$ can be set
proportional to the first two hydrogen energy levels
$E=-1/[2a_{B}n^{2}], n=1,2$. In the last equation,
$a_{B}=\hbar^{2}/me^{2}=1$ is the Bohr radius, which is used as the
length unit in this paper. It is assumed that the band parameters
$t_{v}$ and $t_{c}$ are much smaller than
$\varepsilon_{0}^{c}-\varepsilon_{0}^{v}$, i.e. the bands are
non-overlapping.

The valence and conduction band wave functions in the Bloch
representation
\begin{eqnarray}
\psi_{{\bf k}}^{v}({\bf
r})=e^{i{\bf k}{\bf r}}u_{{\bf k}}^{v}({\bf r})
, \label{psiv_Bloch} \\
\psi_{{\bf k}}^{c}({\bf r})=e^{i{\bf k}{\bf r}}u_{{\bf k}}^{c}({\bf
r}), \label{psic_Bloch}
\end{eqnarray}
can be represented as linear combinations of Wannier functions
$w^{v}({\bf r})$ and $w^{c}({\bf r})$. In fact, one can use for
spatially periodic Bloch functions $u_{{\bf k}}^{v}({\bf r})$ and
$u_{{\bf k}}^{v}({\bf r})$ the following representation:
\begin{eqnarray}
u_{{\bf k}}^{v}({\bf r})=\sum_{{\bf L}}e^{i{\bf k}({\bf r}-{\bf L})}
w^{v}({\bf r}-{\bf L}) \label{psiv}
\end{eqnarray}
and similar for $u_{{\bf k}}^{c}({\bf r})$, where ${\bf
L}=a_{0}(n_{x},n_{y},n_{z})$ are the cell vectors.

Unfortunately, it is difficult to find the exact expressions for the
Wannier functions, and hence for the Bloch functions, even for such
a simple three-dimensional system. For simplicity, we choose the
Wannier functions to be equal to the $1s^{0}$ and $2p^{0}$ hydrogen
wave functions:
\begin{eqnarray}
w^{v}({\bf r})&=&\frac{1}{\sqrt{\pi}} e^{-r}
, \label{wv} \\
w^{c}({\bf r})&=&\frac{1}{4\sqrt{2\pi}} e^{-r/2}z. \label{wc}
\end{eqnarray}
This choice of the Wannier functions is an approximation, since the
orthogonality condition on different sites $\int d{\bf r}w^{l*}({\bf
r}-{\bf L}_{1}) w^{m}({\bf r}-{\bf L}_{2})=\delta^{lm}\delta_{{\bf
L}_{1}{\bf L}_{2}}$ is violated. However, when the lattice parameter
$a_{0}$ is much larger than the Bohr radius, $a_{B}<<a_{0}$, the
requirement of orthogonality is satisfied with a very high
precision, since the overlap between different sites becomes
negligible. In our calculations, we shall use the lattice constant
values $a_{0}=10$ and $20$.

In the next section, we shall study the solution of
Eqs.~(\ref{rhovv}) and (\ref{rhovc}) when an external short pulse
field
\begin{eqnarray}
{\bf E}(t)={\bf E}_{0}e^{-t^{2}/\tau^{2}} \label{electricfield}
\end{eqnarray}
with $\tau\sim 1-10{\rm fs}$ is applied. This will allow us to find
the optical absorption spectrum $A(\omega )$ of the system, which is
defined by the real part of the ratio of the Fourier transforms of
the total polarization
\begin{eqnarray}
P(\omega )=i(4\pi
/\sqrt{\epsilon_{b}})(\varepsilon_{0}^{c}-\varepsilon_{0}^{v})|d^{cv}|^{2}a_{0}^{3}\int
\frac{d{\bf k}}{(2\pi )^3}\int dt e^{i\omega t}\rho_{{\bf k}}^{vc}
(t) \label{P}
\end{eqnarray}
 and the pulse field
$E(\omega )$:
\begin{equation}
A(\omega)=-2{\rm Re}[P(\omega )/E(\omega)] .
 \label{Absorption}
\end{equation}

\section{Results and discussion}
\label{Results}

\subsection{Hartree-Fock}
\label{Hartree-Focksection}

Before proceeding with the solution of the TDDFT equations
(\ref{rhovv}), (\ref{rhovc}) for different xc potentials, we briefly
discuss the case of the Hartree-Fock potential for the two-band
model presented above. Hartree-Fock is the most widely used
approximation in the SBE theory\cite{Haug} for the study of optical
absorption spectra in bulk systems and heterostructures subject to
both smooth and pulsed external fields (see, for example,
Refs.~\onlinecite{Maslov1,Maslov2}).

In the time-dependent Hartree-Fock approximation, the total electron
wave function satisfies the following equation:
\begin{eqnarray}
i\frac{\partial \Psi_{{\bf k}}^{v}({\bf r},t)}{\partial t}
=&~&\left[ -\frac{{\bf \nabla}^{2}}{2}-{\bf E}(t){\bf r}+\int d{\bf
r}'\frac{\sum_{{\bf q}}\Psi_{{\bf q}}^{v*}({\bf r}',t)\Psi_{{\bf
q}}^{v}({\bf r}',t)}{|{\bf r}-{\bf r}'|}\right] \Psi_{{\bf
k}}^{v}({\bf r},t)
\nonumber \\
&-&\int d{\bf r}'\frac{\sum_{{\bf q}}\Psi_{{\bf q}}^{v*}({\bf
r}',t)\Psi_{{\bf q}}^{v}({\bf r},t)}{|{\bf r}-{\bf r}'|} \Psi_{{\bf
k}}^{v}({\bf r}',t). \label{HF}
\end{eqnarray}
Applying the analysis presented in Section \ref{Formalism} [see
Eqs.~(\ref{psi})-(\ref{nrho})], it is possible to show that
Eq.~(\ref{HF}) is equivalent to the system of TDDFT density-matrix
equations (\ref{rhovv}), (\ref{rhovc}) with corresponding matrix
elements for the Hartree and Hartree-Fock potentials. Namely, the
Hartree potential matrix elements are:
\begin{eqnarray}
V_{H{\bf k}}^{lm}(t)=\sum_{n,s=v,c;{\bf q}}I_{H}^{lmns}({\bf k},{\bf
q})\rho_{{\bf q}}^{sn}(t), \label{VH}
\end{eqnarray}
where
\begin{eqnarray}
I_{H}^{lmns}({\bf k},{\bf q})=\int d{\bf r}\int d{\bf r}' \psi_{{\bf
k}}^{l*}({\bf r})\psi_{{\bf k}}^{m}({\bf r}) \frac{\psi_{{\bf
q}}^{n*}({\bf r}')\psi_{{\bf q}}^{s}({\bf r}')}{|{\bf r}-{\bf r}'|}.
\label{IH}
\end{eqnarray}
From Eqs.~(\ref{VH}), (\ref{IH}), one can find the following
approximate expression for the Hartree matrix elements:
\begin{eqnarray}
V_{H{\bf k}}^{lm}(t)= \delta^{lm}V({\bf G}\rightarrow {\bf 0}) +
\sum_{n,s=v,c;{\bf q}} A^{lmns}a_{0}^{3}\int \frac{d{\bf q}}{(2\pi
)^{3}}\rho_{{\bf q}}^{sn}(t)+..., \label{VH2}
\end{eqnarray}
where $V({\bf q})$ is the Fourier transform of the Coulomb potential
$1/|{\bf r}-{\bf r}'|$, ${\bf G}$ is a reciprocal lattice vector and
\begin{eqnarray}
A^{lmns}=\sum_{{\bf G}\not= {\bf 0}}V({\bf G}) \int d{\bf r}
w^{l*}({\bf r})e^{-i{\bf G}{\bf r}}w^{m}({\bf r}) \int d{\bf r}'
w^{n}({\bf r}')e^{i{\bf G}{\bf r}'}w^{s}({\bf r}') \label{A}
\end{eqnarray}
(details of a similar analysis can be found in
Ref.~\onlinecite{Schafer}). In Eq.~(\ref{VH2}), the dots correspond
to the terms proportional to integrals over products of Wannier
functions which reside on different sites [$\sim \int d{\bf r}
w^{l*}({\bf r})w^{m}({\bf r}-{\bf L})$]. These terms are negligible
due to a small overlap of the Wannier functions on different sites
[when $a_{B}\ll a_{0}$, see Eqs.~(\ref{wv}) and (\ref{wc})].
Moreover, the presence of oscillating functions $\exp [-i{\bf G}{\bf
r}]$ and $\exp [i{\bf G}{\bf r}']$ under the integrals in the
expression for the coefficients $A^{lmns}$, Eq.~(\ref{A}), makes the
terms proportional to $A^{lmns}$ much smaller than the first term in
Eq.~(\ref{VH2}). Therefore, the second term in Eq.~(\ref{VH2}) can
be also neglected. Substitution of the remaining term
$\delta^{lm}V({\bf G}\rightarrow {\bf 0})$ into
Eqs.~({\ref{rhovv}}),(\ref{rhovc}) simply causes a constant shift of
the overall potential and can be ignored. Therefore, the Hartree
term does not contribute to the density-matrix equations
({\ref{rhovv}}),(\ref{rhovc}) in this approximation, and we shall
ignore it from now on. It is possible to show \cite{Schafer} that
higher order terms in Eq.~(\ref{VH2}), which we neglect, lead to a
small renormalization of the energy bands and the electric field
time dependence.

In a similar way, one can find an approximate expression for the
matrix elements of the nonlocal Fock exchange potential:
\begin{eqnarray}
V_{HF{\bf k}}^{lm}(t)=-a_{0}^{3}\int \frac{d{\bf q}}{(2\pi
)^{3}}V({\bf k}-{\bf q})\rho_{{\bf q}}^{lm}(t). \label{VHF}
\end{eqnarray}
Substitution of the expression (\ref{VHF}) into Eqs.~(\ref{rhovv}),
(\ref{rhovc}) leads to the following density-matrix equations in the
Hartree-Fock approximation:
\begin{eqnarray}
\frac{\partial}{\partial t}\rho_{{\bf k}}^{vv}(t) = &-&2{\rm
Im}\left[ \left( E(t)d_{\bf k}^{cv}+a_{0}^{3}\int \frac{d{\bf
q}}{(2\pi )^{3}}V({\bf k}-{\bf q})\rho_{{\bf q}}^{cv}(t)\right)
\rho_{{\bf k}}^{vc}(t)\right] , \label{rhovvBloch}
\\
\frac{\partial}{\partial t}\rho_{{\bf k}}^{vc}(t)=
&-&i[\varepsilon_{\bf k}^{v}-\varepsilon_{\bf k}^{c} -a_{0}^{3}\int
\frac{d{\bf q}}{(2\pi )^{3}}V({\bf k}-{\bf q})(\rho_{{\bf
q}}^{vv}(t)-\rho_{{\bf q}}^{cc}(t)) ]\rho_{{\bf k}}^{vc}(t)
\nonumber \\
&-&i [\rho_{{\bf k}}^{cc}(t)-\rho_{{\bf k}}^{vv}(t)] \left(
E(t)d_{\bf k}^{vc}-a_{0}^{3}\int \frac{d{\bf q}}{(2\pi)^{3}}V({\bf
k}-{\bf q})\rho_{{\bf q}}^{vc}(t)\right) . \label{rhovcBloch}
\end{eqnarray}
These equations are equivalent to the standard SBEs.\cite{Haug}) The
solution of the system of Eqs.~(\ref{rhovvBloch}),
(\ref{rhovcBloch}) in the presence of an external field ${\bf E}(t)$
allows one to calculate the optical polarization in the system by
means of Eq.~(\ref{Absorption}).

\begin{figure}[h]
\centering{
\includegraphics[width=8.0cm]{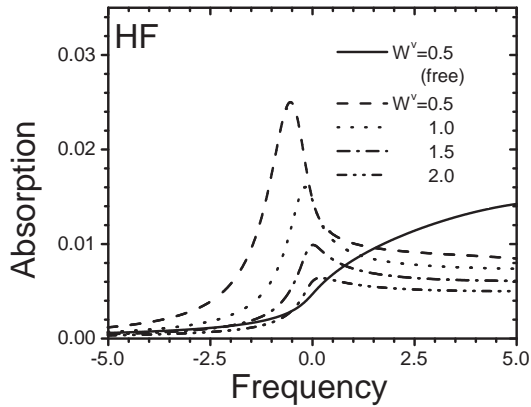}}
\caption{Optical absorption spectra in Hartree-Fock approximation at
different values of the valence bandwidth (divided by $\pi^{2}$).
Here and in the Figures below, we consider external pulses
Eq.~(\ref{electricfield}) of duration $\tau$. For this Figure,
$\tau=0.05$. The conduction bandwidth and the band gap are
$W^{c}=0.4\pi^{2}$ and $\omega_{g}=5$. The amplitude of the electric
field is $E_{0}=2.0$. We consider the case of parabolic bands.}
\label{fig:1}
\end{figure}

In Fig.~\ref{fig:1}, we present the frequency dependence of the
optical absorption spectra at different values of the amplitude of
the external field in Hartree-Fock approximation, when an external
short-time electric pulse Eq.~(\ref{electricfield}) is applied.
Here and in other Figures, we introduce screening $\lambda=1$ in the
Coulomb kernel $1/({\bf q}^{2}+\lambda^{2})$ and in order to reach a
steady state more quickly we introduced a decoherence terms $\Gamma
(1+\alpha |{\bf k}|/\pi ) \rho^{lm}$ on the right hand side of the
density matrix equations. Here $\Gamma =0.2,\alpha =7.5$. In this
paper, in the numerical results for the absorption spectra we set
the prefactor $(8\pi
/\sqrt{\epsilon_{b}})(\varepsilon_{0}^{c}-\varepsilon_{0}^{v})|d^{cv}|^{2}$
to be equal 1 [see Eqs.~(\ref{P}) and (\ref{Absorption})]. The
absorption spectra demonstrate a pronounced excitonic feature with a
shape and a peak position that depend on the external field
amplitude parameters.

The SBE approach was successfully applied to study optical
properties of different bulk materials in the case of weak external
fields, when linear response theory can be applied. Recently, this
formalism was used to study the nonlinear response of different
quantum-well systems,\cite{Maslov1,Maslov2} and some other effects,
like the sideband generation, Franz-Keldysh effect, four-wave
mixing, and higher correlation effects, in particular, biexcitons
etc (see, for example Ref.~\onlinecite{Chemla}). As discussed in the
Introduction, the main shortcoming of the SBE approach for small
(nano-) systems and short pulses is that the Coulomb interaction
effects are treated the mean-field theory level. All the fluctuation
effects are hidden in the inverse scattering (decoherence) time
parameters $\Gamma_{{\bf k}}^{lm}$ , which are usually introduced
into the system of the SBEs (\ref{rhovvBloch}), (\ref{rhovcBloch})
by means of the terms $\Gamma_{{\bf k}}^{lm}\rho_{{\bf k}}^{lm}$.
However, the characteristic times in this case are shorter than the
Coulomb scattering time; therefore, fluctuation effects are very
important and can not be neglected. It is extremely difficult to
include the higher Coulomb terms in the SBEs in a controllable way,
since in this case one needs to consider additional equations for
higher order correlation functions. The TDDFT approach allows one to
treat the Coulomb effects in principle exactly, which should make
this approach favorable in the case of small systems and short-time
pulses, provided one can find a suitable xc functional to account
for correlations. In the following two Subsections, we shall
consider the optical properties of the two-band insulator by means
of the TDDFT formalism developed in Section \ref{Formalism}. We
limit ourselves here to exchange-only functionals.

\subsection{Adiabatic LDA and GGA}
\label{LDAsection}

The LDA is a standard approximation used in DFT to study the ground
state properties of many materials. In the time-dependent case,
there exists a generalization of this approximation - the adiabatic
LDA (ALDA). In the adiabatic approximation it is assumed that the xc
potential at time $t$ depends on the time-dependent particle density
$n({\bf r},t)$ at the same time $t$ only. In other words, all memory
effects are neglected, and the TDDFT equations remain local in time,
which allows one to solve the problem numerically relatively easily.
In ALDA, the x-only potential matrix elements have the following
momentum and time dependence:
\begin{eqnarray}
V_{xLDA{\bf k}}^{lm}(t)=-\left(\frac{3}{\pi}
\right)^{1/3}\int_{cell} d{\bf r} \psi_{{\bf k}}^{l*}({\bf r})
n^{1/3}({\bf r},t) \psi_{{\bf k}}^{m}({\bf r}).
\label{VLDAapproximation}
\end{eqnarray}

\begin{figure}[h]
\centering{
\includegraphics[width=8.0cm]{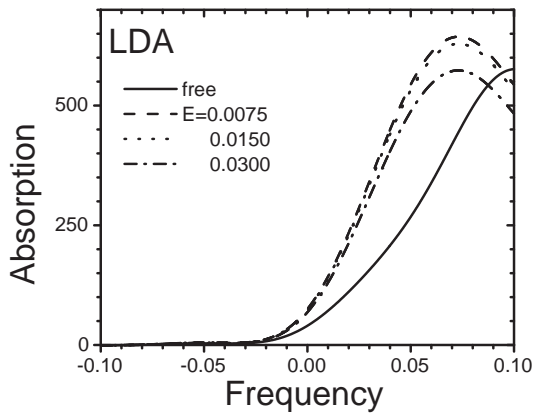}}
\caption{Absorption spectra in ALDA at different values of the
electric field amplitude and the free electron bands $W^{v}=0.15,
W^{c}=0.06$. The other parameters are $\omega_{g}=0.375, a_{0}=20,
\tau =10 ,\Gamma =0.0025$  and $\alpha =7.5$.} \label{fig:2}
\end{figure}

Our numerical analysis confirms the previous observation (see, e.g.
Ref.~\onlinecite{Onida}) that the optical absorption spectra do not
demonstrate excitonic features with the ALDA exchange potential
(\ref{VLDAapproximation}) at various values of the model parameters
(Fig.~\ref{fig:2}). Actually, there is a very wide peak in the
absorption spectra at rather large values of the quasiparticle
masses (or very narrow energy bands); however, this wide absorption
spectra weight cannot be related to a well-defined excitonic energy
value, and hence it does not correspond to an exciton formation. For
details of the numerical calculations in ALDA and other potentials
used in this paper (see below), we refer the reader to
Appendix~\ref{appendix_density}.

Similar results can be obtained for standard adiabatic GGAs, for
example the van Leeuwen-Baerends\cite{vanLeeuwen} (LB) potential. In
the LB approximation, the xc potential has the following form:
\begin{eqnarray}
V_{LB}({\bf r},t)=V_{LDA}({\bf r},t)-\beta n^{1/3}({\bf r},t)
\frac{x^{2}}{1+3\beta x\ln (x+\sqrt{x^{2}+1})}, \ \ \ \
x=\frac{|\nabla n({\bf r},t)|}{n^{4/3}({\bf r},t)},
\label{VLBapproximation}
\end{eqnarray}
where $\beta$ is a parameter, which depends on the material. This
potential has an important feature for finite systems, namely, a
correct ($\sim 1/r$) asymptotic behavior at long distances (The ALDA
potential decreases exponentially at $r\rightarrow\infty$). In spite
of this, no excitonic features are produced.

Thus, we have seen that the standard adiabatic LDA+GGAs for the xc
potentials does not allow one to reproduce the excitonic features in
the optical absorption spectra within the TDDFT formalism developed
above. One needs to go beyond this approximation and to find another
class of simple TDDFT potentials which demonstrate excitonic peaks
in the absorption spectra. As shown by Kim and G\" orling
\cite{Kim1,Kim2} such potentials exist. The main requirement is a
$1/q^{2}$ singularity in the exchange energy kernel. In the
following Subsection, we shall consider the optical absorption
spectra for classes of functionals where the underlying exchange
energy has such a singularity.

\subsection{The Slater and KLI potentials}
\label{SlaterKLIsection}

As shown above, the ALDA cannot describe excitonic effects in the
optical absorption spectra. A fundamental shortcoming of the LDA
approximation is that it contains a self-interaction error. In order
to reduce this error, several approximations for self-interaction
corrected (SIC) potentials, were proposed. Probably the most often
used SIC potential is due to Perdew and Zunger.\cite{Perdew}
However, in their scheme one must deal with orbital-dependent
potentials, which makes the calculations difficult, especially in
the time-dependent case. To make a computational procedure simpler,
the method of the optimized effective potential (OEP) was proposed
(for overview and references, see for example Ref.~\cite{Ullrich}).
In this approximation, all the wave functions for different orbitals
satisfy a Schroedinger equation with a common, orbital-independent,
xc potential $v_{xc\sigma}^{OEP}({\bf r})$. A time-dependent
generalization of the OEP approximation was given in
Refs.~\onlinecite{Ullrich2,Gross}. Unfortunately, this method is
computationally very demanding in the time-dependent case. Krieger,
Li, and Iafrate\cite{Krieger} proposed a simplified method (the KLI
scheme) for the OEP in the equilibrium exact-exchange case with
$v_{xc\sigma}^{OEP}({\bf r})$ depending explicitly on the orbital
functions $\varphi_{j\sigma}$. In the equilibrium case, the KLI
potential is defined by the following integral equation:
\begin{eqnarray}
v_{xc\sigma}^{KLI}({\bf r})=\sum_{j}\frac{n_{j\sigma}({\bf
r})}{n_{\sigma}({\bf r})}\left\{ u_{xcj\sigma ({\bf r})} + \int
d^{3}r' |\varphi_{j\sigma} ({\bf r}')|^{2} [v_{xc\sigma}^{KLI}({\bf
r}')-u_{xcj\sigma ({\bf r}')}]\right\}, \label{VKLI}
\end{eqnarray}
were the orbital-dependent potentials $u_{xcj\sigma}({\bf r})$ are
obtained from the xc energy $E_{xc}$ as follows:
\begin{eqnarray}
u_{xcj\sigma}({\bf r})=[f_{j\sigma}\varphi_{j\sigma}^{*}({\bf
r})]^{-1}\delta E_{xc}[\{\varphi_{j\sigma}\}]/\delta
\varphi_{j\sigma}({\bf r}). \label{uKLI}
\end{eqnarray}
This approximation was generalized to the time dependent case in the
adiabatic approximation ($n({\bf r})\rightarrow n({\bf r},t)$) in
Ref.~\onlinecite{Ullrich2}.

In the linear regime, the optical absorption spectra of insulators
show excitonic peaks, when an OEP approximation with the exchange
energy with a kernel that contains a $1/q^{2}$ singularity is
used.\cite{Kim1,Kim2} However, it is much more difficult to use this
approach for short and strong pulses, when one needs to go beyond
linear response.

Here, we consider simplified OEPs with a Coulomb singularity in the
exchange energy kernel, which are approximate solutions of
Eqs.~(\ref{VKLI}), (\ref{uKLI}), and solve the problem in the
time-domain by using the TDDFT formalism developed in the previous
Sections. This approach can be applied for external fields of
arbitrary strength and duration.

\begin{figure}[h]
\centering{
\includegraphics[width=8.0cm]{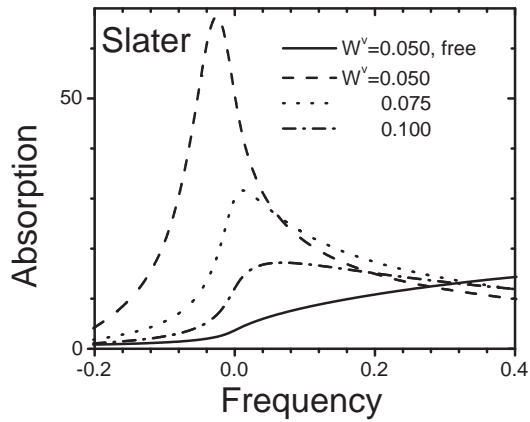}}
\caption{Optical absorption spectra in Slater approximation at
different values of the valence bandwidth (divided by $\pi^{2}$).
The other parameters are $W^{c}=0.025\pi^{2}, \omega_{g}=0.2,
a_{0}=10,\tau=4, E=0.0166,\Gamma =0.012$ and $\alpha =7.5$.}
\label{Slater}
\end{figure}

\begin{figure}[h]
\centering{
\includegraphics[width=8.0cm]{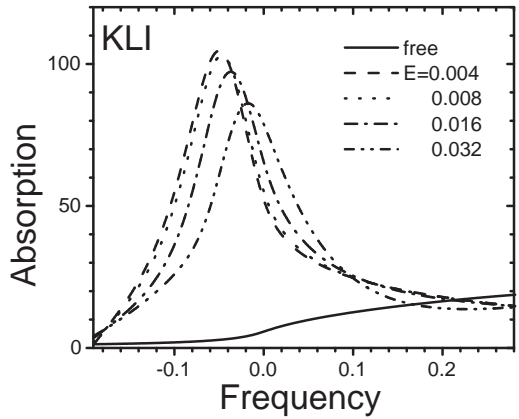}}
\caption{Optical absorption spectra in KLI approximation at
different values of the electric field. The other parameters are
$W^{v}=0.0416\pi^{2}, W^{c}=0.0208\pi^{2}, \omega_{g}=0.166,
a_{0}=10,\tau=4.8,\Gamma =0.012,$ and $\alpha =7.5$.} \label{KLI}
\end{figure}

We consider the KLI potential and  the Slater potential
\cite{Krieger}, which can be obtained if one neglects the
orbital-dependent constants on the right hand side of
Eq.~(\ref{VKLI}):
\begin{eqnarray}
v_{xc\sigma}^{Slater}({\bf r},t)=\sum_{j}\frac{n_{j\sigma}({\bf
r},t)}{n_{\sigma}({\bf r},t)}u_{xcj\sigma ({\bf r},t)}.
\label{VSlater}
\end{eqnarray}
We use the Fock exchange energy for our two-band model in
Eq.~(\ref{uKLI}). Optical absorption spectra obtained with the
Slater and KLI exchange potentials at different values of the model
parameters are presented in Figs.~\ref{Slater} and \ref{KLI}.

As can be seen from these Figures, the absorption spectra
demonstrate a significant excitonic peak. In particular, the
excitonic peak amplitude and the exciton binding energy decrease
with increasing valence bandwidth in the Slater case
(Fig.~\ref{Slater}). This effect takes place because the increase of
the bandwidth is equivalent to a reduction of the effective mass of
electron-hole pairs. A similar behavior can be found in the case of
the KLI potential. Fig.~\ref{KLI} shows the KLI absorption spectrum
at different values of the electric field. Here, the spectrum
demonstrates nonlinear effects when the pulse amplitude is large.

\begin{figure}[h]
\centering{
\includegraphics[width=8.0cm]{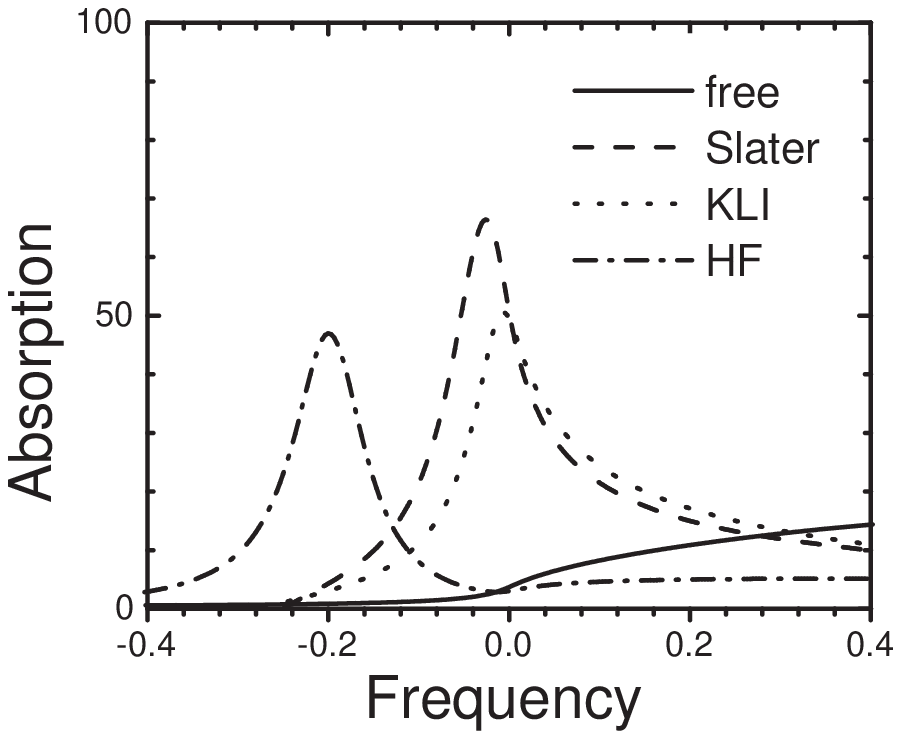}}
\caption{Optical absorption spectra in different approximations at
the same values of the model parameters.  The valence bandwidth is
equal to $0.05\pi^{2}$ and the other parameters are given in the
caption to Fig.~\ref{Slater}. The Hartree-Fock result is obtained
using a Bloch function approximation for the xc potential matrix
elements (see Appendix).} \label{comparison}
\end{figure}

There is a significant quantitative difference between different xc
potentials. In particular, we find that the excitonic features
become more pronounced as one passes from KLI to Slater. The
excitonic effects are defined mainly by the ratio of the Coulomb
interaction energy to the typical energy of the free system. Since
in our case the valence and conduction bands and the gap have the
same order of magnitude, we can consider the ratio of the Coulomb
energy to the valence bandwidth. In the Slater approximation, the
amplitude of the Coulomb interaction energy appears to be bigger
compared to KLI. In fact, as follows from the Appendix, the Coulomb
interaction matrix elements are much larger in the Slater
approximation (see Eqs.~(\ref{A_approximation}), (\ref{B1}) and
(\ref{B2})). In the strongly localized case, considered in this
paper, the Slater results are closer to HF (Fig.~\ref{comparison}).
For less localized electrons, when $a_{0}\sim 1$ and the momentum
dependence of the xc potential matrix elements can not be neglected,
we expect to get closer agreement between Slater and KLI. Another
important conclusion from our numerical results is a decrease of the
excitonic binding energy in Slater and KLI compared to HF. In fact,
it is known that the HF approximation gives an overestimated binding
energy of excitons (for a discussion, see
Ref.~\onlinecite{Bruneval}).

Thus, we have shown that TDDFT can describe excitonic effects in the
optical absorption spectra triggered by short-time laser pulses,
when the exchange energy kernel contains a long-range Coulomb
singularity.

\section{Conclusions}
\label{Conclusions}

In this paper, a density-matrix TDDFT formalism to describe
ultrafast processes in semiconductor structures has been developed.
We have shown that the corresponding system of equations for the
matrix elements is a generalization of the SBEs, since it allows one
to take into account Coulomb effects in principle exactly. This is
very important in the case of short pulses and when the system is
very small, such as for nanostructures. Another important advantage
of this formalism is that it can be applied directly in the time
domain, which makes the calculation much faster. As an example, we
studied the optical absorption spectra of a two-band model bulk
insulator. We have shown that the optical absorption spectra include
excitonic effects when the xc energy kernel contains a $1/q^{2}$
singularity. The exciton binding energy within the Slater and KLI
approaches is much smaller compared to HF, which often gives too
large value for the binding energy.

This TDDFT approach should be useful for studying ultrafast
processes in real semiconductor and polymer nanostructures, for
example for short-time laser pulse experiments. In particular, it
gives access to a variety of nonlinear effects, which will be the
subject of future studies.

\section*{Acknowledgements}
We thank Paul de Boeij for useful discussions. We acknowledge
financial support from NSF under Grant No. DMR-0553485.

\appendix

\section{Technical details}
\label{appendix_density}

In this Appendix, we present some useful approximations and
technical details of the numerical solution of the TDDFT equations
(\ref{rhovv}), (\ref{rhovc}). Since, generally speaking, evaluation
of the xc matrix elements beyond LDA requires six-dimensional
spatial integration, it is necessary to make some simplifications to
speed up the calculations. We shall use the Wannier and Bloch
representation (\ref{psiv_Bloch})-(\ref{wc}) for the wave functions.
In this case, the electron density can be expressed in the following
form:
\begin{eqnarray}
n({\bf r},t)=2\sum_{l,m,{\bf L},{\bf q}} \rho_{{\bf q}}^{lm}(t)
e^{-i{\bf L}{\bf q}}w^{m*}({\bf r}-{\bf L}) w^{l}({\bf r}).
\label{density}
\end{eqnarray}
Since the wave functions $w^{m}({\bf r})$ and $w^{l}({\bf r})$ are
strongly localized, the main contribution to expression
(\ref{density}) is given by the term ${\bf L}={\bf 0}$. Therefore,
the electron density is approximately
\begin{eqnarray}
n({\bf r},t)\simeq 2\sum_{l,m} \rho_{tot}^{lm}(t) w^{m*}({\bf r})
w^{l}({\bf r}), \label{density2}
\end{eqnarray}
where
\begin{eqnarray}
\rho_{tot}^{lm}(t)=\sum_{{\bf q}} \rho_{{\bf q}}^{lm}(t).
\label{rhotot}
\end{eqnarray}

It turns out that Eq.~(\ref{density2}) can be obtained by another
approximation for the wave functions, which follows from
Eqs.~(\ref{psiv_Bloch})-(\ref{psiv}). Namely,
\begin{eqnarray}
\psi_{{\bf k}}^{v}({\bf r})\simeq e^{i{\bf k}{\bf r}}u_{{\bf
k}_{0}}^{v}({\bf r})
, \label{psiv_Bloch_approx} \\
\psi_{{\bf k}}^{c}({\bf r})\simeq e^{i{\bf k}{\bf r}}u_{{\bf
k}_{0}}^{c}({\bf r}). \label{psic_Bloch_approx}
\end{eqnarray}
We choose the wave vector ${\bf k}_{0}$ to be zero. This
approximation is good when one considers excitation processes around
the direct gap. In this case,
\begin{eqnarray}
\psi_{{\bf k}}^{v}({\bf r})\simeq e^{i{\bf k}{\bf r}} w^{v}({\bf r}), \label{psiv_approx} \\
\psi_{{\bf k}}^{c}({\bf r})\simeq e^{i{\bf k}{\bf r}} w^{c}({\bf
r}). \label{psic_approx}
\end{eqnarray}
By using this approximation one gets the following expression for
the time-dependent matrix elements of the LDA and Slater potentials:
\begin{eqnarray}
V_{LDA{\bf k}}^{lm}(t)\simeq -\left(\frac{3}{\pi} \right)^{1/3}
\int_{cell} d{\bf r} w^{l*}({\bf r}) w^{m}({\bf r})\left( \sum_{a,b}
\rho_{tot}^{ba}(t) w^{a*}({\bf r}) w^{b}({\bf r})\right)^{1/3},
\label{LDA_approximation}
\end{eqnarray}
\begin{eqnarray}
V_{Slater{\bf k}}^{lm}(t)\simeq -\sum_{n,s,{\bar n},{\bar
s}}\sum_{{\bf p},{\bf q}} V({\bf p}-{\bf q})A^{lmns{\bar n}{\bar
s}}(t)\rho_{{\bf p}}^{sn}(t) \rho_{{\bf q}}^{{\bar s}{\bar n}}(t),
\label{Slater_approximation}
\end{eqnarray}
where
\begin{eqnarray}
A^{lmns{\bar n}{\bar s}}(t)\simeq \int_{cell} d{\bf r}
\frac{w^{l*}({\bf r}) w^{m}({\bf r})w^{n*}({\bf r}) w^{s}({\bf
r})w^{{\bar n}*}({\bf r}) w^{{\bar s}}({\bf r})}{n({\bf r},t)}
\label{A_approximation}
\end{eqnarray}
and the density $n({\bf r},t)$ is given in Eq.~(\ref{density2}).

Finally, by using the Wannier wave functions (\ref{psiv})-(\ref{wc})
and making an approximation similar to that used in the derivation
of Eqs.~(\ref{LDA_approximation})-(\ref{Slater_approximation}), one
gets the following equation for the KLI matrix elements, which
follows from Eq.~(\ref{VKLI}):
\begin{eqnarray}
V_{KLI{\bf k}}^{lm}(t)=V_{Slater{\bf k}}^{lm}(t)+\sum_{n,s,{\bar
n},{\bar s},{\bf p}} B_{1}^{lmns}(t)\rho_{{\bf p}}^{sn} V_{KLI{\bf
p}}^{{\bar n}{\bar s}}(t)\rho_{{\bf p}}^{{\bar s}{\bar n}}
+\sum_{n,{\bar n},s,{\bar s},{\bf p},{\bf q}} B_{2{\bf
p}}^{lmns{\bar n}{\bar s}}(t) V({\bf p}-{\bf q})\rho_{{\bf q}}^{s
n}\rho_{{\bf p}}^{{\bar s}{\bar n}},\nonumber \\
\label{VKLI_approximation2}
\end{eqnarray}
where ${\bar V}_{Slater{\bf p}}(t)$ is defined in
Eqs.~(\ref{Slater_approximation}) and (\ref{A_approximation}), and
\begin{eqnarray}
B_{1}^{lmns}(t)= \int_{cell} d{\bf r} \frac{w^{l*}({\bf r})
w^{m}({\bf r})w^{n*}({\bf r}) w^{s}({\bf r})}{n({\bf r},t)},
\label{B1}
\end{eqnarray}
\begin{eqnarray}
B_{2{\bf p}}^{lmns{\bar n}{\bar s}}(t)&\simeq& -\int_{cell} d{\bf r}
\frac{w^{l*}({\bf r}) w^{m}({\bf r})w^{n*}({\bf r}) w^{s}({\bf r})
w^{{\bar n}*}({\bf r})w^{{\bar s}}({\bf r})}{n({\bf r},t)}
\sum_{a,b}w^{a*}({\bf r})w^{b}({\bf r}) \rho_{{\bf
p}}^{ba} \nonumber \\
&\simeq& -\int_{cell} d{\bf r} \frac{w^{l*}({\bf r}) w^{m}({\bf
r})w^{n*}({\bf r}) w^{s}({\bf r}) w^{{\bar n}*}({\bf r})w^{{\bar
s}}({\bf r})w^{v*}({\bf r})w^{v}({\bf r})}{n({\bf r},t)}. \label{B2}
\end{eqnarray}
It is convenient to solve Eq.~(\ref{VKLI_approximation2}) by
iteration. Therefore, we have reduced the six-dimensional space
integral to a three-dimensional one. For numerical integration, we
divide the space interval into 120 parts for every direction. In the
case of momentum integration, we divide the interval into 100 parts.
As it was mentioned in Section \ref{Hartree-Focksection}, we use a
decoherence factor $\Gamma$ in equations in order to reach a steady
state faster. We have found that the position of the excitonic peak
does not depend on value of $\Gamma$, but its height decreases with
$\Gamma$ increasing. Also, it was found that in the Slater and KLI
cases the total absorption conservation law is satisfied with a
higher precision when $\Gamma$ increases. Due to finite values of
the time interval and time step, absorption $A(\omega)$ acquires
unphysical finite weights at large values of $|\omega -\omega_g|$.
These weights disappear when $\Gamma$ increases.

For completeness, let us present approximate matrix elements for the
Hartree-Fock potential derived within the same approximation. These
expressions are necessary to make a comparison of the absorption
spectra using different potentials (Fig.~\ref{comparison}). The
Hartree-Fock matrix elements have the following structure:
\begin{eqnarray}
V_{HF{\bf k}}^{lm}(t)\simeq -\sum_{n,s}\sum_{{\bf p},{\bf q}} V({\bf
k}-{\bf p})A^{lmns}\rho_{{\bf p}}^{sn}(t), \label{HF_approximation}
\end{eqnarray}
where
\begin{eqnarray}
A^{lmns}\simeq \int_{cell} d{\bf r} w^{l*}({\bf r}) w^{m}({\bf
r})w^{n*}({\bf r}) w^{s}({\bf r})\label{HF_approximation}.
\end{eqnarray}
Following the discussion in Sec.~\ref{Hartree-Focksection}, the
quantities (\ref{HF_approximation}) can be approximated as
$A^{lmns}\simeq\delta^{ls}\delta^{mn}$. However, in approximation
(\ref{psiv_Bloch_approx}),(\ref{psic_Bloch_approx}) these elements
become
\begin{eqnarray}
A^{lmns}\simeq \delta^{mn}I^{m}\int_{cell} d{\bf r} w^{l*}({\bf r})
w^{s}({\bf
r})\simeq\delta^{mn}\delta^{ls}I^{m}\label{HF_approximation2},
\end{eqnarray}
where $I^{m}, m=v,c,$ is the main contribution of the function
$|w^{m}({\bf r})|^{2}$ to the integral Eq.~(\ref{HF_approximation}).
Its value can be chosen to be equal to the maximum value of
$|w^{m}({\bf r})|^{2}$, i.e. $1/\pi$ for the valence band function
and $1/(16\pi e^{2})$ for the conduction band function. In the last
case, there is an additional factor $1/2$, which comes from the
azimuthal angle integration. The factor $\pi/6$ must be added in
Eq.~(\ref{HF_approximation}), since in Slater and KLI there is an
additional momentum integration over the Brillouin zone, and the
integration is performed over a sphere of the radius $\pi$ instead
of the cubic Brillouin zone.

\begin{figure}[h]
\centering{
\includegraphics[width=8.0cm]{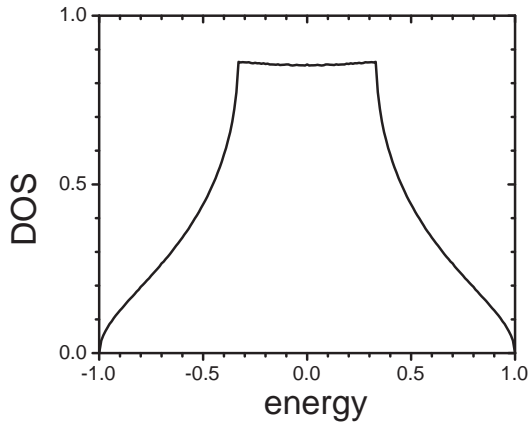}}
\caption{ The three-dimensional density of states
Eq.~(\ref{densityofstates}) as a function of energy.}
\label{fig_densityofstates}
\end{figure}

The most important property of expressions
(\ref{LDA_approximation})-(\ref{Slater_approximation}) and
(\ref{VKLI_approximation2}) is the fact that they are
momentum-independent. Therefore, as follows from Eqs.~(\ref{rhovv})
and (\ref{rhovc}), the matrix elements $\rho_{{\bf q}}^{ml}(t)$ are
functions of the energy
\begin{eqnarray}
\varepsilon_{{\bf k}} = \frac{1}{3}[\cos k_{x}+\cos k_{y}+\cos
k_{z}]. \label{ek}
\end{eqnarray}
In order to reduce the dimensionality of the integrals it is
convenient to introduce the corresponding three-dimensional density
of states
\begin{eqnarray}
D(\varepsilon ) &=&\int \frac{d^{3}p}{(2\pi )^{3}}\delta
[\varepsilon -(1/3)(\cos q_{x}+\cos q_{y}+\cos q_{z})] \nonumber \\
&=& \frac{3}{\pi^{3}}\int_{0}^{\pi}d p_{x}\int_{0}^{\pi}d p_{y}
\frac{\theta [1-(3\varepsilon -\cos p_{x}-\cos
p_{y})^{2}]}{\sqrt{1-(3\varepsilon -\cos p_{x}-\cos p_{y})^{2}}},
\label{densityofstates}
\end{eqnarray}
where the energy $\varepsilon$ changes from $-1$ to $1$. The energy
dependence of the density of states is shown in
Fig.~\ref{fig_densityofstates}.

Since the xc matrix elements depend on integrals which have the
structure
\begin{eqnarray}
\int \frac{d^{3}p}{(2\pi )^{3}}\int \frac{d^{3}q}{(2\pi )^{3}}
V({\bf p}-{\bf q})F(\rho_{{\bf p}}^{lm},\rho_{{\bf q}}^{ns}),
\label{densityofstates2}
\end{eqnarray}
it is also convenient to introduce a two-energy density of states
\begin{eqnarray}
D_{2} (\varepsilon ,{\bar \varepsilon}) =\int \frac{d^{3}p}{(2\pi
)^{3}}\int \frac{d^{3}q}{(2\pi )^{3}} &~&\delta [\varepsilon
-(1/3)(\cos p_{x}+\cos p_{y}+\cos p_{z})] \nonumber \\
&~&\times\delta [{\bar \varepsilon} -(1/3)(\cos q_{x}+\cos
q_{y}+\cos q_{z})] V({\bf p}-{\bf q}). \label{densityofstates2_2}
\end{eqnarray}
Performing the integration over $p_{z}$ and $q_{z}$ in
Eq.~(\ref{densityofstates2_2}), one finds
\begin{eqnarray}
D_{2} (\varepsilon ,{\bar \varepsilon})
=\left(\frac{3}{\pi}\right)^{2}\int \frac{d^{2}p_{||}}{(2\pi
)^{2}}\int \frac{d^{2}q_{||}}{(2\pi )^{2}}&~&\frac{\theta
[1-x^{2}]}{\sqrt{1+\delta-x^{2}}} \frac{\theta
[1-y^{2}]}{\sqrt{1+\delta-y^{2}}} \nonumber \\
&\times&\frac{1}{ ({\bf p}_{\parallel } -{\bf q}_{\parallel})^{2}
+(\cos^{-1}x -\cos^{-1}y)^{2}
+1/\lambda^{2} },\nonumber \\
\label{densityofstates2_3}
\end{eqnarray}
where $x=3\varepsilon -\cos p_{x}-\cos p_{y}, y=3{\bar \varepsilon}
-\cos q_{x}-\cos q_{y}$ and ${\bf p}_{\parallel }=(p_{x},p_{y},0)$.
In Eq.~(\ref{densityofstates2_3}), we have introduced "screening"
parameters $\delta$ and $\lambda$, since in the three-dimensional
case the function $D_{2} (\varepsilon ,{\bar \varepsilon})$ has a
logarithmic singularity at $\varepsilon ={\bar \varepsilon}$. In
fact, it is possible to show analytically that in the case of a
square dispersion $D_{2} (\varepsilon \rightarrow {\bar
\varepsilon}) \sim \ln (1/|\sqrt{\varepsilon}-\sqrt{{\bar
\varepsilon}}|)$. We apply this approximation in the LDA case
(Fig.~(\ref{fig:2})). In other Figures, we use the parabolic band
approximation for sake of simplicity. This approximation gives
results close to those obtained by using the densities of states
Eqs.~(\ref{densityofstates}) and (\ref{densityofstates2_3}), when
one uses a decoherence factor with a momentum cutoff (see Section
\ref{Hartree-Focksection}). In fact, in this case low momenta give a
dominant contribution to the absorption processes.

\end{document}